\documentclass[twocolumn,noshowpacs]{revtex4}
\usepackage{amsmath,graphicx,amsbsy,amssymb,amsmath,epsfig,latexsym,wasysym,pifont,mathrsfs,natbib}
\usepackage{float}
\newfloat{widefig}{thp}{lop}
\usepackage{comment}
\usepackage{verbatim}
\usepackage{multirow,array}
\usepackage{hyperref}

\begin{document}

\title{First principles study of the structural phase stability and magnetic order in various structural phases of Mn$_2$FeGa}

\author{Ashis Kundu}
\email[]{k.ashis@iitg.ernet.in}
\author{Subhradip Ghosh}
\email[]{subhra@iitg.ernet.in}

\affiliation{Department of Physics, Indian Institute of Technology Guwahati, Guwahati-781039, Assam, India.}

\date{\today}
 
\begin{abstract} 
We investigate the structural and magnetic properties of Mn$_{2}$FeGa for different phases(cubic, hexagonal and tetragonal) reported experimentally using density functional theory. The relative structural stabilities, and the possible phase transformation mechanisms are discussed using results for total energy, electronic structure and elastic constants. We find that the phase transformation form  hexagonal to ground state tetragonal structure would take place through a Heusler-like phase which has a pronounced electronic instability. The electronic structures, the elastic constants and the supplementary phonon dispersions indicate that the transition from the Heusler-like to the tetragonal phase is of pure Jahn-Teller origin. We also describe the ground state magentic structures in each phase by computations of the exchange interactions. For Heusler-like and tetragonal phases, the ferromagnetic exchange interactions associated with the Fe atoms balance the dominating antiferromagnetic interactions between the Mn atoms leading to collinear magnetic structures. In the hexagonal phase, the direction of atomic moment are completely in the planes with a collinear like structure, in stark contrast to the well known non-collinear magnetic structure in the hexagonal phase of Mn$_{3}$Ga, another material with similar structural properties. The overwhelmingly large exchange interactions of Fe with other magnetic atoms destroy the possibility of magnetic frustration in the hexagonal phase of Mn$_{2}$FeGa. This comprehensive study provides significant insights into the microscopic physics associated with the structural and magnetic orders in this compound.
\end{abstract}

\maketitle

\section{INTRODUCTION}
Mn$_{2}$YZ compounds in Heusler and Heusler-like structures have been in the attention of spintronics community due to their multiple possible applications in spin-transfer torque random access memory(STT-MRAM)~\cite{WinterlikAM12}, spin valves~\cite{SinghPRL13}, magnetic tunnel junction~\cite{KubotaAPL09}, spin gapless semiconductor~\cite{OuardiPRL13} and magnetic shape memory effect~\cite{LiuAPL05}. Such traits of Mn$_{2}$YZ compounds are artefacts of various possible structures in which the systems can crystallise as well as of different orientations of the Mn spins giving rise to interesting magnetic configurations~\cite{LiuAPL05,BarmanPRB08,PaulJAP11,MeshcheriakovaPRL14,WinterlikPRB11,GasiAPL13,
WollmannPRB14,WollmannPRB15,NayakPRL13,Kundumodulation17}. Among the compounds in this family, Mn$_{3}$Ga~\cite{KurtPSS11,RodePRB13,KharelJPCM14,ZhangJPCM13,KhmelevskyiPRB16} and Mn$_{2}$NiGa~\cite{LiuAPL05,Kundumodulation17} have been studied extensively. Mn$_{2}$NiGa is a recently discovered ferrimagnetic shape memory system which martensitically transforms to a low temperature tetragonal structure through a series of complex modulated phases of orthorhombic and monoclinic symmetries~\cite{Kundumodulation17}; the martensitic transformation being driven by phonon softening and Fermi surface nesting~\cite{PaulJPCM15}. Mn$_{3}$Ga, on the other hand, is found to crystallise in a cubic Cu$_{3}$Au-like~\cite{KharelJPCM14}, a hexagonal DO$_{19}$~\cite{KurtAPL12} and a tetragonal DO$_{22}$ phase~\cite{RodePRB13, KurtPRB11}, resulting in a high perpendicular magneto-crystalline anisotropy in the DO$_{22}$ phase and a large Exchange Bias in the DO$_{19}$ phase, which are useful for STT-MRAM~\cite{RodePRB13} and magnetic tunnel junction~\cite{KurtAPL12} applications, respectively. First-principles Density Functional Theory (DFT) calculations predicted a Heusler-like metastable structure in Mn$_{3}$Ga with a half-metallic gap, which phase transforms to the DO$_{22}$ structure~\cite{WollmannPRB14, WollmannPRB15}. This phase, however, has not yet been synthesised experimentally. Comprehensive DFT calculations inferred that the phase transition from DO$_{19}$ to DO$_{22}$, as observed in the experiments~\cite{KharelJPCM14}, happens via the Heusler-like phase~\cite{ZhangJPCM13}. First-principles computations of the magnetic exchange interactions~\cite{KhmelevskyiPRB16} and the magnetic anisotropy~\cite{RodePRB13} concluded that the novel magnetic properties of Mn$_{3}$Ga in DO$_{19}$ and DO$_{22}$ structures are due to non-collinear magnetic structures arising out of frustrations due to geometry as well as competing exchange interactions between in-plane and out-of-plane Mn atoms.

Inspite of enough promises towards a variety of magnetism related applications, both Mn$_{3}$Ga and Mn$_{2}$NiGa  have low saturation magnetisations originating from the predominantly antiferromagnetic interactions between the Mn atoms. This hinders the exploitations of their complete potentials in the respective applications. The low saturation magnetisation in Mn$_{3}$Ga limits it's applicability in potential permanent magnet applications, inspite of having strong uniaxial magnetic anisotropy and large Curie temperature. For Mn$_{2}$NiGa, much lower saturation magnetisation~\cite{LiuAPL05,PaulJAP11}, in comparison to Ni$_{2}$MnGa, the prototype magnetic shape memory system in the Heusler family, is an obstacle to obtain significant magnetic field induced strain required for actuator applications. Attempts have, therefore, been made to combat the dominant antiferromagnetic interactions in these systems by replacing one of the Mn with ferromagnetic elements like Co and Fe. Complete replacement of one Mn atom by Fe, resulting in the compound Mn$_{2}$FeGa, was expected to circumvent the problems. Synthesis of Mn$_{2}$FeGa, however, exhibited several important aspects of structure-property relationship. Various crystallographic phases with possible magnetic structures were observed and predicted experimentally. Gasi {\it et al.}~\cite{GasiAPL13} reported an inverse tetragonal Heusler structure with a low saturation magnetic moment and a high Curie temperature for samples annealed at 400$^{0}$C. The system behaved like an exchange spring which was attributed to the distribution of Fe atoms among the Mn sites resulting in two different magnetic states of Fe. A Cu$_{3}$Au-like structure was observed when the system was annealed at 800$^{0}$C. A giant tuneable Exchange Bias was later obtained in the inverse tetragonal structure with a saturation magnetisation as low as 0.09 $\mu_{B}$/f.u.~\cite{NayakNM15}  suggesting a near compensation of moments from Mn and Fe atoms. DFT calculations, like Mn$_{3}$Ga, predicted a meta-stable inverse Heusler phase for Mn$_{2}$FeGa, which transforms to the inverse tetragonal structure observed in the experiments~\cite{WollmannPRB14,WollmannPRB15}. The electronic structure of inverse Heusler Mn$_{2}$FeGa revealed a pronounced instability associated with the minority spin band~\cite{WollmannPRB15}. However, unlike Mn$_{2}$NiGa, the transformation from inverse Heusler to inverse tetragonal phase didn't turn out to be volume conserving, implying the possibility of absence of shape memory effect in this system. Investigations into sputter deposited thin films of Mn$_{2}$FeGa~\cite{NiesenIEEE16} find that the system crystallises either in inverse tetragonal or in the Cu$_{3}$Au-like cubic phase. Mossbauer spectroscopy on epitaxial thin films suggested that Fe might be statistically distributed among the two Mn sites leading to a low spin moment in the inverse tetragonal phase~\cite{BettoarXiv17}. Heteroepitaxially grown thin films of Mn-Fe-Ga with composition near that of Mn$_{2}$FeGa also exhibited strong perpendicular anisotropy and moderate coercivity suggesting that this system can be used in mid-range permanent magnet applications~\cite{KalachearXiv17}. Like Mn$_{3}$Ga, this system could also be synthesised in DO$_{19}$-type hexagonal structure which yielded a giant Exchange Bias field upto 1.32 kOe~\cite{LiuAPL16}. The large Exchange Bias was attributed to the presence of substantial ferromagnetic matrix due to Fe-Mn pairs in an antiferromagnetic host.

The experimental observations clearly indicate that investigations into the phase stability, sub-lattice occupancy and magnetic properties of Mn$_{2}$FeGa would be insightful. The following issues, in particular, are worth looking into, in order to understand the physics associated with Fe substitution in either Mn$_{3}$Ga or in Mn$_{2}$NiGa, the two compounds having very different functional aspects: (i) the energetics of various phases observed experimentally or are predicted theoretically but not observed experimentally, (ii) the magnetic interactions among three different magnetic atoms and their dependencies on the sub-lattice occupancies of the atoms in different structural phases. This can be particularly relevant in the hexagonal phase where the Fe atom has more than one choice of the crystallographic site it can occupy, and (iii) the effect of anti-site disorder on the magnetic properties. In this paper, we have undertaken a DFT based comprehensive investigation into the structural and magnetic properties of the four crystallographic phases of Mn$_{2}$FeGa, the Cu$_{3}$Au-like, the hexagonal DO$_{19}$-like, the inverse Heusler and the inverse tetragonal. Emphasis has been given on identifying the magnetic structure in the hexagonal phase as it is supposed to be most complex in this phase as was observed in Mn$_{3}$Ga. We have computed the magnetic exchange interactions in each structural phase in order to understand the magnetic structures. The electronic structures of the structural phases are computed and analysed in order to provide a possible picture of the phase transformations. The elastic constants and the phonon dispersion relations for select phases are computed to supplement the analysis from the energetics and the electronic structures. In this work, we have not incorporated anti-site disorder in order to avoid dealing with prohibitively large number of possible configurations, in particular for the hexagonal structure. In absence of concrete quantitative estimate of the anti-site disorder from the experiments, this is justified. 
The paper is organised as follows: in the section II, computational details are given. The results and discussion are presented in section III followed by conclusions.

\section{Calculational details}
All calculations were performed with spin-polarised density functional theory (DFT) based projector augmented wave(PAW) method~\cite{PAW94} as implemented in Vienna Ab-initio Simulation Package (VASP)~\cite{VASP196,VASP299}. Perdew-Burke-Ernzerhof (PBE96) implementation of Generalised Gradient Approximation (GGA) for exchange and correlation~\cite{PBEGGA96} part in the Hamiltonian was used throughout. An energy cut-off of 500 eV and a Monkhorst-Pack~\cite{MP89} $25 \times 25 \times25$ $k$-mesh for Cu$_{3}$Au-like and inverse Heusler structures , a $15 \times 15 \times13$ $k$-mesh for the tetragonal structure and a $13 \times 13 \times 11$ $k$-mesh for the hexagonal structure were used for self consistent calculations. Larger $k$-meshes were used for the densities of states calculations of all the structures. For all calculations, the total energy convergence criteria   and the force convergence criteria were set to  10$^{-6}$ eV and  to $10^{-2}$ eV/\r{A}  respectively. The elastic constants were calculated from the second derivatives of the total energies with respect to the strain tensors~\cite{Vitosbook07}.

The magnetic pair exchange parameters were calculated with multiple scattering Green function formalism as implemented in SPRKKR code~\cite{EbertRPP11}. In here, the spin part of the Hamiltonian is mapped to a Heisenberg model

\begin{eqnarray}
H = -\sum_{\mu,\nu}\sum_{i,j}
J^{\mu\nu}_{ij}
\mathbf{e}^{\mu}_{i}
.\mathbf{e}^{\nu}_{j}
\end{eqnarray}
$\mu$, $\nu$ represent different sub-lattices,
\emph{i}, \emph{j} represent atomic positions and $\mathbf{e}^{\mu}_{i}$ denotes the unit vector
along the direction of magnetic moments at site \emph{i} belonging to sub-lattice $\mu$. The $J^{\mu \nu}_{ij}$s are calculated from the energy differences due to infinitesimally small orientations of a pair of spins within the formulation of Liechtenstein et al.~\cite{LiechtensteinJMMM87}. In order to calculate the energy differences, full potential spin polarised scaler relativistic Hamiltonian with angular momentum cut-off $l_{max} = 3$ is used along with a converged $k$-mesh for Brillouin zone integrations. The Green's functions are calculated for 32 complex energy points distributed on a semi-circular contour. The energy convergence criterion is set to 10$^{-5}$ eV for the self-consistency. The equilibrium lattice parameters and optimised atomic positions as obtained from the PAW calculations are used to obtain the self-consistent potentials in the multiple scattering Green's function method. 

\section{Results and Discussions}

\begin{figure}[t]
\centerline{\hfill
\psfig{file=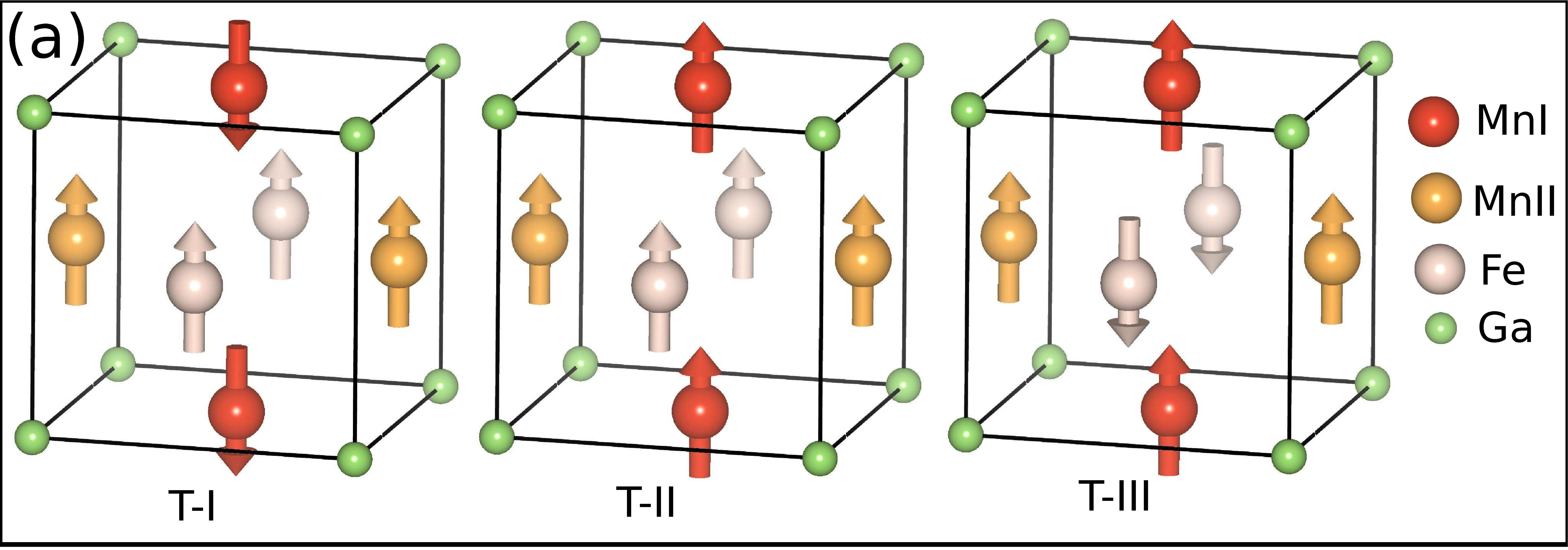,width=0.4\textwidth}\hfill}
\vspace{0.5 cm}
\centerline{\hfill
\psfig{file=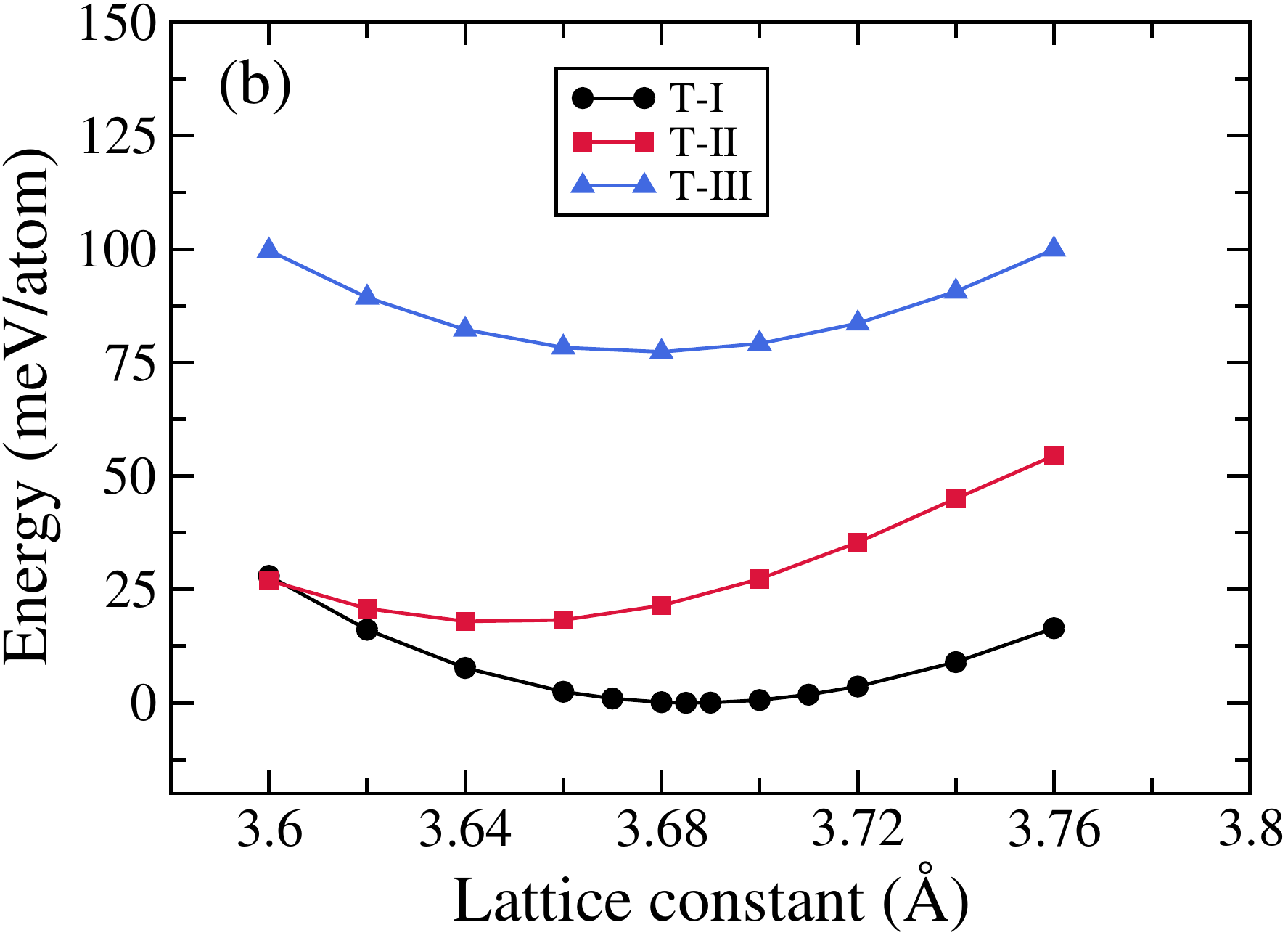,width=0.35\textwidth}\hfill}
\caption{(a) Possible magnetic configurations in Cu$_{3}$Au phase of Mn$_{2}$FeGa. (b) Total energy as a function of lattice constant for different magnetic configurations in Cu$_{3}$Au phase of Mn$_{2}$FeGa.}
\label{fig1}
\end{figure}

\subsection{Structural parameters and magnetic structures in various crystallographic phases}
Experimentally, Mn$_{2}$FeGa has been observed to crystallise in three different phases, the Cu$_{3}$Au-like, the tetragonal DO$_{22}$ like and the hexagonal DO$_{19}$~\cite{GasiAPL13,LiuAPL16}, depending on the annealing temperature. DFT calculations predicted a inverse Heusler phase~\cite{WollmannPRB14,WollmannPRB15,LuoJAP08}, yet undetected in experiments.  In these investigations, the possible magnetic structures associated with different structural phases have been indicated through indirect evidences. In this sub-section, we present results on the structural properties and possible ground state magnetic structures, along with atomic moments for all four structural phases of Mn$_{2}$FeGa. Throughout the manuscript, the Cu$_{3}$Au-like phase, the hexagonal phase, the inverse Heusler phase and the inverse tetragonal phases are referred to as the Cu$_{3}$Au, the DO$_{19}$, the X$_{a}$ and the L1$_{0}$ phases respectively. The results are summarised in  Table \ref{table1}.

\begin{table*}[ht]
\begin{center}
\caption{\label{table1} The calculated lattice parameters(in \AA), total(M) and atomic magnetic moments(M$_{X}$)(in $\mu_{B}$ per formula unit) of Mn$_{2}$FeGa in Cu$_{3}$Au, X$_{a}$, L1$_{0}$ and  DO$_{19}$ phases. The results from experiments and other DFT calculations are given in parenthesis}
\begin{tabular}
          {l@{\hspace{0.5cm}}  ll@{\hspace{0.67cm}}  ll@{\hspace{0.67cm}}  ll@{\hspace{0.5cm}} ll@{\hspace{0.5cm}}  ll@{\hspace{0.5cm}}  ll@{\hspace{0.5cm}} ll@{\hspace{1.0cm}}}
\hline\hline
Structure         &Lattice parameter $a$ &Lattice parameter $c$  &  M  &  M$_{\rm MnI}$  & M$_{\rm MnII}$ & M$_{\rm Fe}$ &  M$_{\rm Ga}$ \\ \hline
Cu$_{3}$Au     &3.69(3.7526~\cite{GasiAPL13}) &3.69  &  0.68   &   -3.08  &   2.55  &  1.28   &  -0.03  \\
               
X$_{a}$        &5.78(5.80~\cite{LuoJAP08},5.79~\cite{WollmannPRB14}) &5.78 &  1.04(1.03~\cite{WollmannPRB14})   &  -2.07   &   2.80  &  0.29   &  0.01  \\
               
L1$_{0}$     & 3.68(3.7915~\cite{GasiAPL13},3.68~\cite{WollmannPRB15}) &7.29(7.1858~\cite{GasiAPL13},7.29~\cite{WollmannPRB15}) &  -0.80(0.96~\cite{WinterlikAM12},-0.78~\cite{WollmannPRB15})   &  -2.43   &  2.96   &   -1.41  &  0.04  \\
                   
DO$_{19}$ &5.25(5.3151~\cite{LiuAPL16}) &4.20(4.3050~\cite{LiuAPL16})  &  1.22(1.26~\cite{LiuAPL16})   &   -2.84  &   2.54  &  1.55   &  -0.03  \\
                  
\hline\hline
\end{tabular}
\end{center}
\end{table*}

\textbf{Cu$_{3}$Au phase:} In the ordered Cu$_{3}$Au structure of Mn$_{2}$FeGa, the Ga atom occupies the corners of the cubic unit cell while the two Mn, MnI and MnII, along with the Fe, occupy the face centres (Fig. \ref{fig1}(a)). Depending upon the relative orientations of the moments of the three magnetic atoms,  three inequivalent collinear spin configurations are possible in this structure as shown in Fig.\ref{fig1}(a).  Our total energy calculations show that the energetically lowest configuration is T-I where the two Mn atoms orient anti-parallel (Fig. \ref{fig1}(b)). The calculated lattice parameter agrees well with experimental reported one (Table I). The total magnetic moment  is only 0.68 $\mu_{B}/f.u.$, with MnI having a larger moment among the two Mn atoms. The Fe moment is only 1.28 $\mu_{B}$, much less than that in it's elemental solid phase.  Mossbauer spectra of Mn$_{2}$FeGa, annealed at 800$^{\circ}$C~\cite{GasiAPL13}, confirmed the Cu$_{3}$Au-like structure but the measurements of hyperfine field distributions suggested that the average moment at the Fe site is only $0.5 \mu_{B}$ as Fe may occupy Mn sites as well. Since we have not included the anti-site disorder, that is mixing of Fe and Mn atoms at all crystallographic sites, the discrepancy with the experimental results is natural. Since the experimental measurements did not indicate any possibility of non-collinearity in the magnetic structure, our consideration of the collinear magnetic structure is justified, and thus, our results provide the possible ground state magnetic configuration if Mn$_{2}$FeGa crystallises in a perfectly ordered Cu$_{3}$Au phase. A comparison with Mn$_{3}$Ga in the ordered Cu$_{3}$Au phase~\cite{KharelJPCM14} shows that replacing one Mn atom by Fe has led to a reduction of the total moment of the system. In Mn$_{3}$Ga, the MnI and MnII moments were -2.93 $\mu_{B}$ and 2.24 $\mu_{B}$ respectively, leading to a total moment of 1.55 $\mu_{B}$ per formula unit. The atomic moments in Mn$_{2}$FeGa, as obtained here, suggest that the Mn moments have not changed significantly, but the reduction in the total moment, in comparison to Mn$_{3}$Ga, happens due to weaker exchange splitting associated with the Fe atoms.

\begin{figure}[t]
\centerline{\hfill
\psfig{file=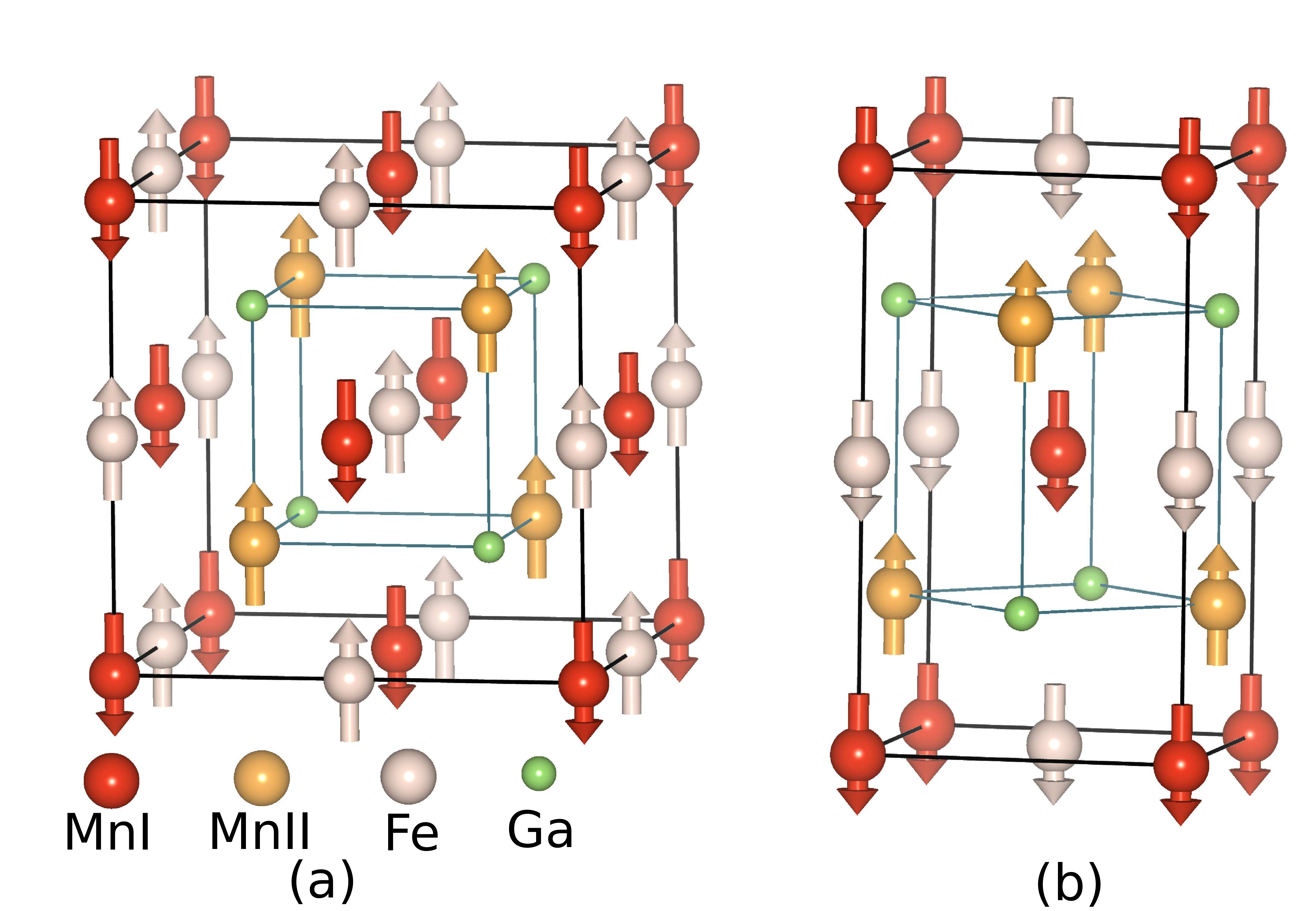,width=0.40\textwidth}\hfill}
\caption{(a) The X$_{a}$ structure and (b) the L1$_{0}$ structure of Mn$_{2}$FeGa. The spin configurations shown are the ground state configurations obtained from our calculations.}
\label{fig2}
\end{figure}

\textbf{X$_{a}$ phase:} DFT calculations predict a Hg$_2{}$CuTi structure (space group no. 216;
$F\bar{4}3m$) for Mn$_{2}$FeGa with four inequivalent Wyckoff positions(4a, 4b, 4c, 4d) in the unit cell~\cite{LuoJAP08,WollmannPRB14}. In this structure, also known as X$_{a}$ structure, the MnI and MnII  atoms occupy 4a(0,0,0) and 4c(0.25, 0.25, 0.25) positions respectively. The 4b(0.5, 0.5, 0.5) and 4d(0.75, 0.75, 0.75) sites are occupied by Fe and Ga atoms respectively. The X$_{a}$ structure of Mn$_{2}$FeGa is shown in Fig. \ref{fig2}(a). The equilibrium lattice constant (Table \ref{table1}) obtained in this work is consistent with the reported results~\cite{LuoJAP08,WollmannPRB14}. The calculated magnetic moment is close to 1.0 $\mu_{B}/f.u$, in very good agreement with the existing DFT results. The  atomic moments (Table \ref{table1}) show that the Fe moment is quenched substantially, and that the low total moment arises due to anti-parallel alignment of the two Mn atoms. The ground state magnetic configuration in this structure is consistent with  that of Mn$_{2}$NiGa~\cite{PaulJAP11}, in which Fe substitution at the Ni site gives rise to the present compound under investigation. Interestingly, the total and the atomic moments of Mn$_{2}$FeGa in this structural phase are very similar to Mn$_{2}$NiGa. In contrast, there is substantial differences between the total and the atomic moments of Mn$_{2}$FeGa and Mn$_{3}$Ga in this phase. In Mn$_{3}$Ga, the total moment is nearly zero in this phase~\cite{WollmannPRB14}. This is due to the fact that unlike Mn$_{2}$FeGa, Mn$_{3}$Ga has a Heusler like coordination (space group 225) and thus the Mn moments at 4a and 4b sites compensate the moment at 4c sites associated with the other Mn aligning anti-parallel. The alteration in the coordination in Mn$_{2}$FeGa, thus, introduces magnetic interactions, very different from those in Mn$_{3}$Ga. 

\textbf{L1$_{0}$ phase:} The tetragonal L$_{10}$ phase of Mn$_{2}$FeGa (Fig. \ref{fig2}(b)) has been synthesised~\cite{GasiAPL13,BettoarXiv17} and been the subject of intense investigations due to the possible PMA and high coercivity in this structure.  Investigations into the magnetic structure concluded that the system has a ferrimagnetic order, and unlike Mn$_{3}$Ga~\cite{RodePRB13}, the spins are collinear. The magnetisation measurement on polycrystalline samples at 10 T field yielded a non-saturating magnetisation of about 1.5 $\mu_{B}$~\cite{GasiAPL13}.  Using results from magnetometry,  Betto {\it et. al.}~\cite{BettoarXiv17} estimated the atomic magnetic moments and the spin structure (Fig. \ref{fig2}(b)). Their estimation of atomic moments, based upon magneto-optical sum rules resulted in equal moments on MnI and Fe sites with MnII sites having a 30$\%$ less moment. Incorporation of Fe anti-sites in their calculations increased the moment at MnII sites resulting in nearly equal moments at all three sites, which would result in a very small net moment in the system. This was in contrast to Mn$_{3}$Ga, where the estimated moments associated with the MnI and MnII were 2.67 $\mu_{B}$ and 4.74 $\mu_{B}$ respectively, producing a net moment of about 0.6 $\mu_{B}$ per formula unit~\cite{RodePRB13}. DFT calculations by Wollmann {\it et al.}~\cite{WollmannPRB15}, on the other hand, obtained unequal magnetic moments at MnI and MnII sites, along with a moment at the Fe site which is nearly 50$\%$ of the moment which is larger among the two Mn moments, producing a net moment of about 0.8 $\mu_{B}$ per formula unit. Our results in Table \ref{table1} are in excellent agreement with those of Wollmann {\it et al}. Particularly interesting is the gain in Fe moment in comparison to the X$_{a}$ phase, and the change in it's orientation with respect to the Mn sub-lattices. Such behaviour of the atomic moments, in particular of Fe in Mn$_{2}$FeGa, is in stark contrast to that of Ni in Mn$_{2}$NiGa. In prototype shape memory alloys like Ni$_{2}$MnGa and in Mn$_{2}$NiGa, the systems undergo martensitic transformations from L2$_{1}$ Heusler and X$_{a}$ structures, respectively to L1$_{0}$ structures, conserving their volumes with little changes in their atomic and total magnetic moments. The very different behaviour of the atomic magnetic moments in X$_{a}$ and L1$_{0}$ phases of Mn$_{2}$FeGa indicate that the physics associated with the system in these phases would be different from those of shape memory Ni$_{2}$MnGa and Mn$_{2}$NiGa.

\begin{figure}[t]
\centerline{\hfill
\psfig{file=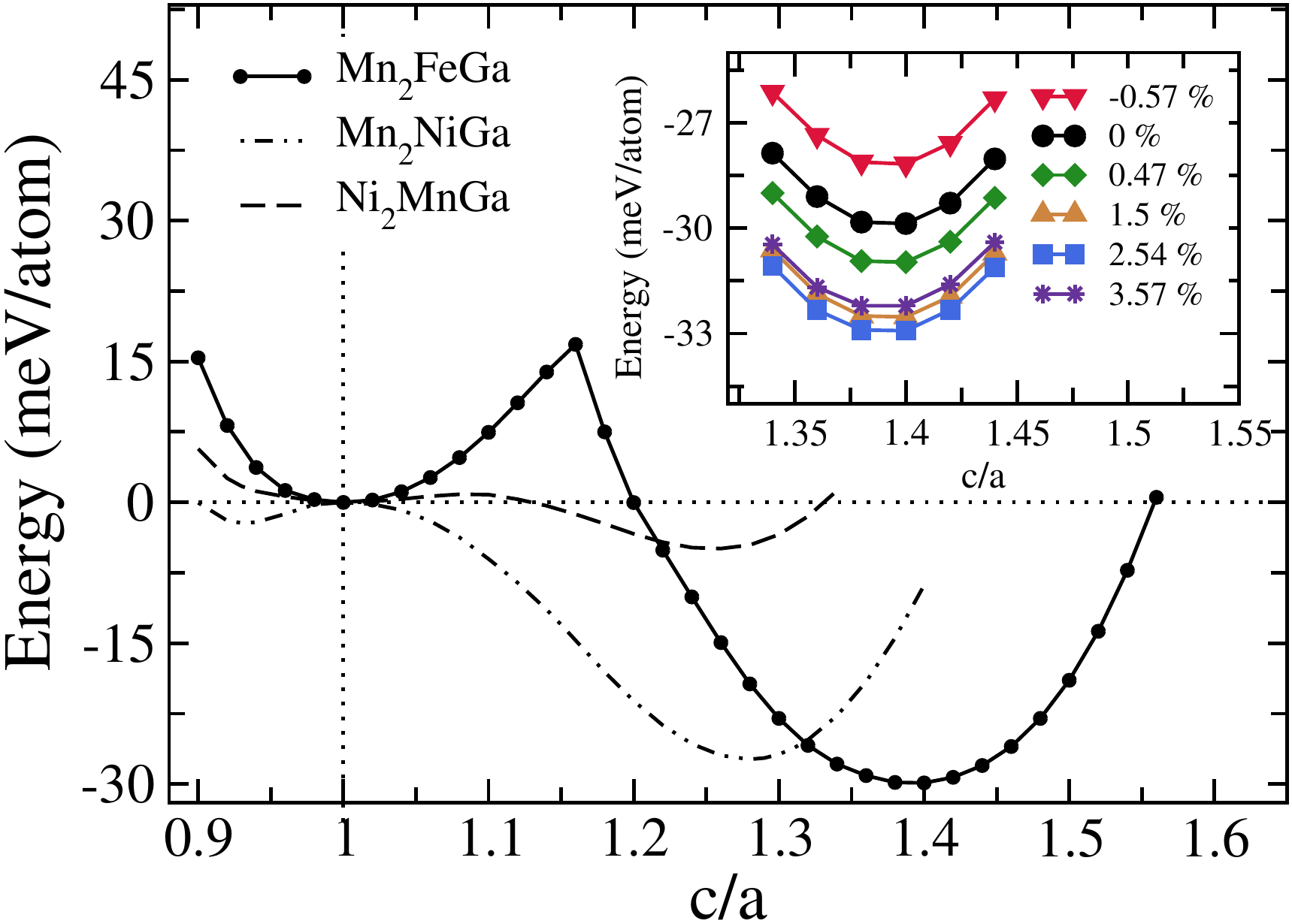,width=0.4\textwidth}\hfill}
\caption{Total energy as a function of tetragonality(c/a) for Mn$_{2}$FeGa in the L1$_{0}$ phase.  The total energy curves of Ni$_{2}$MnGa and Mn$_{2}$NiGa are presented for comparison. The zero energy is taken to be the energy corresponding to the X$_{a}$ phase(c/a=1). Total energy of Mn$_{2}$FeGa as a function of c/a for different volumes in L1$_{0}$ phase are shown in inset. Each curve in the inset represents the percentage change in volume with respect to the equilibrium volume in the X$_{a}$ phase.}
\label{fig3}
\end{figure}

In order to elucidate this point, we have optimised the structural parameters in the L1$_{0}$ phase of Mn$_{2}$FeGa by computing the total energy as a function of the tetragonal deformation, represented by the $c/a$ ratio, keeping the volume fixed at that of the X$_{a}$ phase. The results are presented in Fig. \ref{fig3}. The energetics of Ni$_{2}$MnGa and Mn$_{2}$NiGa are presented for comparison. We find a local energy minima at $c/a = 1.0$ and a global energy minima at $c/a = 1.40$. For Mn$_{2}$FeGa, the energy surface looks very different compared to those of Mn$_{2}$NiGa and Ni$_{2}$MnGa. We further computed the total energies for the L1$_{0}$ structure by varying the volume. The results are shown in the inset of Fig. \ref{fig3}. We clearly see that the optimised volume of the L1$_{0}$ phase is higher by 2.54\% than the volume of X$_{a}$ structure which indicates that the X$_{a}$ to L1$_{0}$ transformation in this system is not a martensitic one and the system would not exhibit shape-memory effect. The optimised structural parameters, presented in Table \ref{table1} are in good agreement with the results obtained from experiments and previous DFT calculations.

\begin{figure}[t]
\centerline{\hfill
\psfig{file=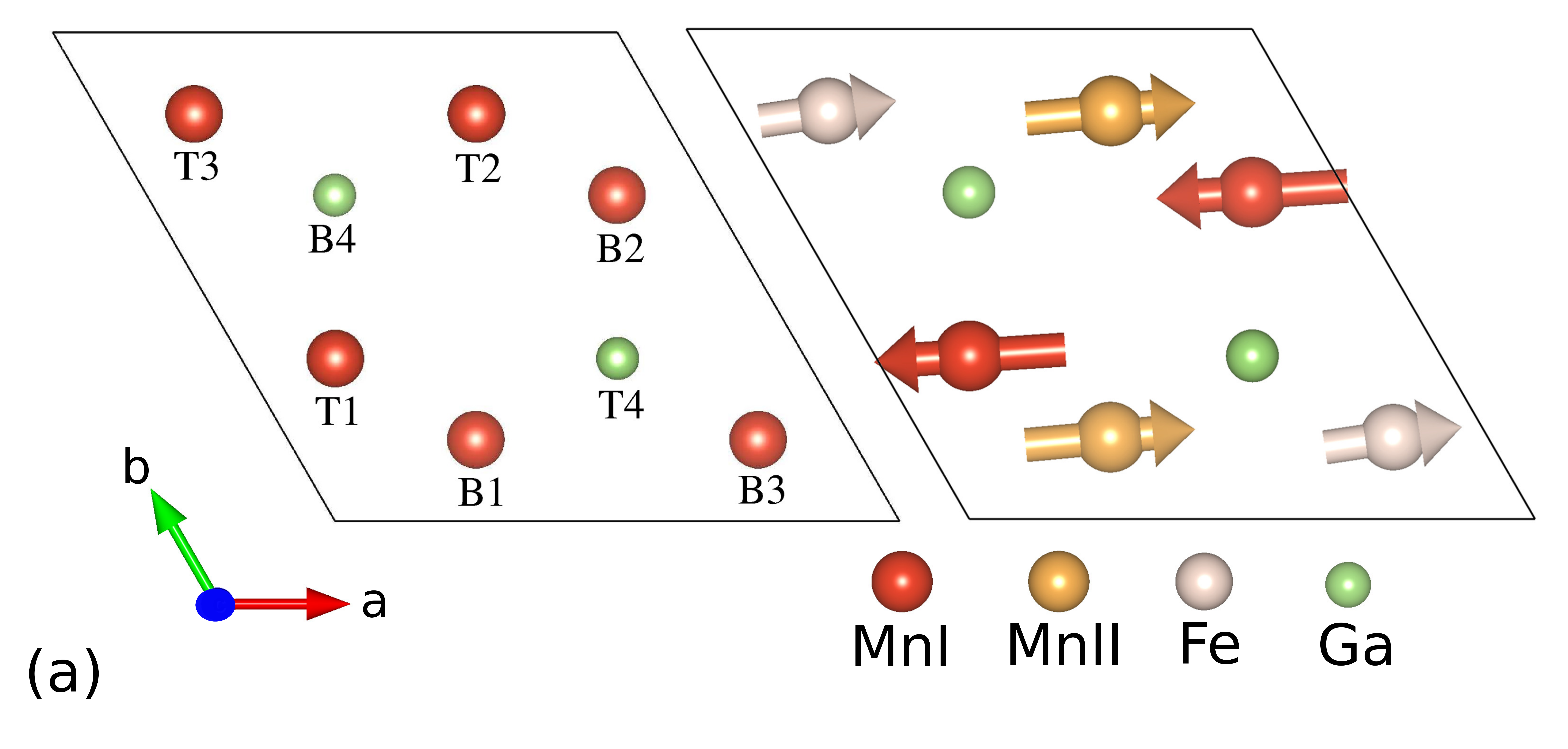,width=0.40\textwidth}\hfill}
\vspace{0.5 cm}
\centerline{\hfill
\psfig{file=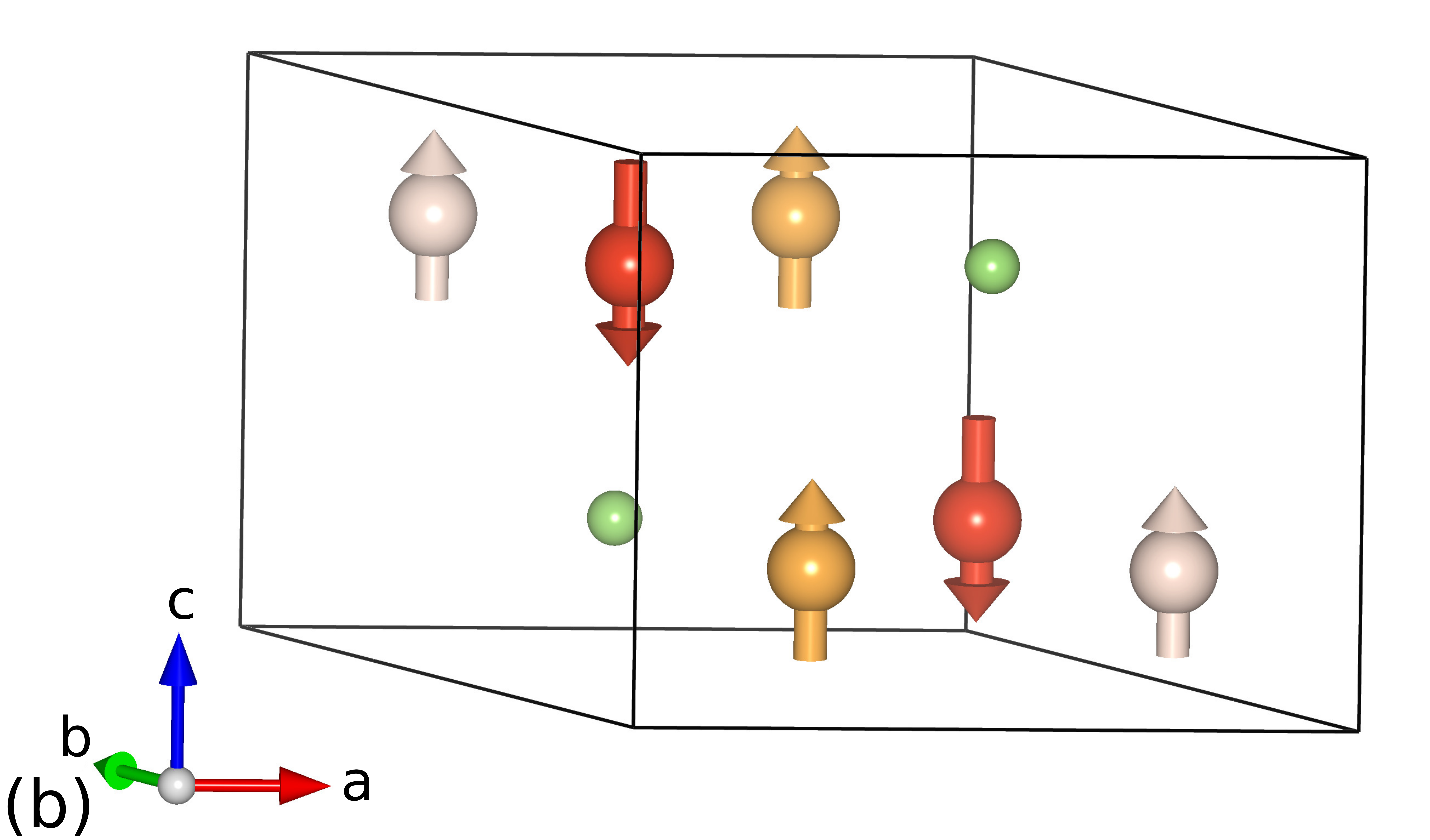,width=0.30\textwidth}\hfill}
\caption{(a) Left: Hexagonal(DO$_{19}$) structure of Mn$_{3}$Ga. The atomic sites  marked T and B correspond to sites in the top and the bottom planes respectively. (a) Right: The DFT calculated magnetic structure of Mn$_{2}$FeGa. (b) The magnetic structure of Mn$_{2}$FeGa in DO$_{19}$ phase modelled after the actual one shown in (a), with the magnetisation axis along $z$-direction.}
\label{fig4}
\end{figure}
\textbf{DO$_{19}$ phase:} The hexagonal DO$_{19}$ phase of  Mn$_{2}$FeGa has been synthesised only recently~\cite{LiuAPL16} after annealing the polycrystalline sample to 600$^{0}$C. 
In the hexagonal(DO$_{19}$) structure(space group no. 194; $P6_{3}/mmm$) there are two formula units of atoms(8 atoms) arranged in two different planes of the unit cell. (T1, T2, T3, T4) and (B1, B2, B3, B4) are the lattice sites in top and bottom planes respectively, as shown in Fig. \ref{fig4}(a). The atomic positions  at top plane are: T1(1/6, 1/3, 3/4), T2(2/3, 5/6, 3/4), T3(1/6, 5/6, 3/4), T4(2/3, 1/3, 3/4),  and at bottom plane are: B1(1/3, 1/6, 1/4), B2(5/6, 2/3, 1/4), B3(5/6, 1/6, 1/4), B4(1/3, 2/3, 1/4). In hexagonal Mn$_{3}$Ga, the Ga atoms occupy T4 and B4 sites in  two different layers of the unit cell. The Mn atoms occupy other six positions in the unit cell. Such a geometry inherently produces a frustrated magnetic structure in Mn$_{3}$Ga. Accordingly, different non-collinear spin configurations were investigated to obtain the ground state magnetic configuration in the system by DFT calculations~\cite{ZhangJPCM13}.  The configuration with the lowest energy was found to be a triangular structure with a 120$^{0}$ mutual orientations of the neighbouring Mn atoms in the same plane as well as of the neighbours in the adjacent planes. This was in confirmation with the results of Neutron diffraction~\cite{krenSSC70}. Subsequently, the magnetic moment came out to be nearly zero with the moment at Mn site to be $\sim$ 2.5 $\mu_{B}$~\cite{krenSSC70,ZhangJPCM13}. The magnetisation measurements on DO$_{19}$ Mn$_{2}$FeGa obtained a saturation moment of 1.26 $\mu_{B}$ per formula unit~\cite{LiuAPL16}. The authors proposed that this is an effect of increasing Fe-Mn local ferromagnetic matrix in otherwise antiferromagnetic host as Fe can occupy Mn sites either in a single layer or across the layers of the hexagonal planes.  If hexagonal Mn$_{2}$FeGa is considered to be formed out of hexagonal Mn$_{3}$Ga by replacing one Mn atom with Fe, in  the DO$_{19}$ unit cell, two Mn atoms out of six are to be replaced by Fe atoms. Since these Fe atoms can occupy any two positions out of the six inequivalent sites among the two planes (keeping the position of Ga atoms fixed), many inequivalent combinations of site occupancy are possible  in this phase. Consideration of symmetry, however, reduces the possible number of configurations to only seven which are given in Table \ref{table2}.  Due to the geometrical frustration arising out of triangular networks in the DO$_{19}$ structure, consideration of non-collinear spin structures as the starting point of the calculations are necessary. In absence of any experimental information regarding the spin structure of this system in DO$_{19}$ phase, we started our calculations with the reported spin configuration of Mn$_{3}$Ga in the literature~\cite{ZhangJPCM13}. We optimised the magnetic structures and calculated the formation enthalpies of all seven configurations.  The results are presented in Table \ref{table2}. The results suggest that the  configuration with configuration number 4 in Table \ref{table2}  has the lowest formation enthalpy.  The corresponding magnetic structure is shown in Fig. \ref{fig4}(a). The calculated total magnetic moment for this configuration is 1.30 $\mu_{B}/f.u.$  which agrees very well with the experimental magnetisation value of 1.26 $\mu_{B}/f.u.$~\cite{LiuAPL16}.

\begin{table}[H]
\centering
\caption{\label{table2} The possible configurations due to different site occupancy patterns  in DO$_{19}$ phase of Mn$_{2}$FeGa. E$_{f}$(eV/f.u.) is the formation enthalpy for a given configuration. M$_{tot}$($\mu_{B}/f.u.$) is the calculated total magnetic moment in a given configuration.}
\begin{tabular}{c@{\hskip 0.1in} c  c  c  c  c  c  c c@{\hskip 0.2in} c@{\hskip 0.2in} c}
\hline \hline
\multicolumn{11}{c}{Sites and occupancies}\\ \hline
No.    &  T1  &  T2  &  T3  &  T4  &  B1  &  B2  &  B3  &  B4  & E$_{f}$ & M$_{tot}$\\ \hline
  1           &  Mn  &  Fe  &  Mn  &  Ga  &  Mn  &  Fe  &  Mn  &  Ga & -0.460 & 3.03\\
  2           &  Mn  &  Fe  &  Mn  &  Ga  &  Mn  &  Mn  &  Fe  &  Ga & -0.459 & 1.63\\
  3           &  Mn  &  Fe  &  Mn  &  Ga  &  Fe  &  Mn  &  Mn  &  Ga & -0.504 & 1.28\\
  4           &  Mn  &  Mn  &  Fe  &  Ga  &  Mn  &  Mn  &  Fe  &  Ga & -0.509 & 1.30\\
  5           &  Mn  &  Fe  &  Fe  &  Ga  &  Mn  &  Mn  &  Mn  &  Ga & -0.456 & 1.46\\
  6           &  Fe  &  Mn  &  Fe  &  Ga  &  Mn  &  Mn  &  Mn  &  Ga & -0.456 & 1.46\\
  7           &  Fe  &  Fe  &  Mn  &  Ga  &  Mn  &  Mn  &  Mn  &  Ga & -0.456 & 1.45\\

                  
\hline\hline
\end{tabular}
\end{table}

Contrary to the expectations, our calculations show the minimum energy configuration of the spins to be one where they are confined to the planes and pointing either along the $a$-direction or against it, a collinear arrangement when observed in the $a-b$ plane. Since the magnetic arrangement is like a collinear one, we calculated the total energy of this configuration by aligning the spins along the $z$-direction. The differences in energies of the two configurations turned out to be less than 1 meV per formula unit. The magnetic moment of the configuration with spins directed either along or against $z$-direction is 1.22 $\mu_{B}$ per formula unit (Table \ref{table1}), very close to that obtained for the minimum energy configuration with spins in the $a-b$ plane (Table \ref{table2}). This justifies using the collinear configuration with $z$-axis the easy axis for further calculations. The lattice parameters obtained by optimising this configuration, presented in Table \ref{table1}, agree quite well with the experiments. Thus, the presence of Fe in the system appears to have taken care of the frustration in the system and produces a collinear kind of magnetic structure which explains the significant non-zero magnetisation of Mn$_{2}$FeGa in this phase. It is worth mentioning that such collinear like magnetic structures are also obtained for some of the other configurations considered here, but they are energetically higher, yet possible to form as is obvious from the negative values of the formation enthalpies (Table \ref{table2}). The optimised magnetic structure corresponding to each of the configurations considered here are presented in the supplementary information.



\subsection{Stabilities of various phases: analysis from energetics, electronic structure and elastic constants}
\begin{figure}[t]
\centerline{\hfill
\psfig{file=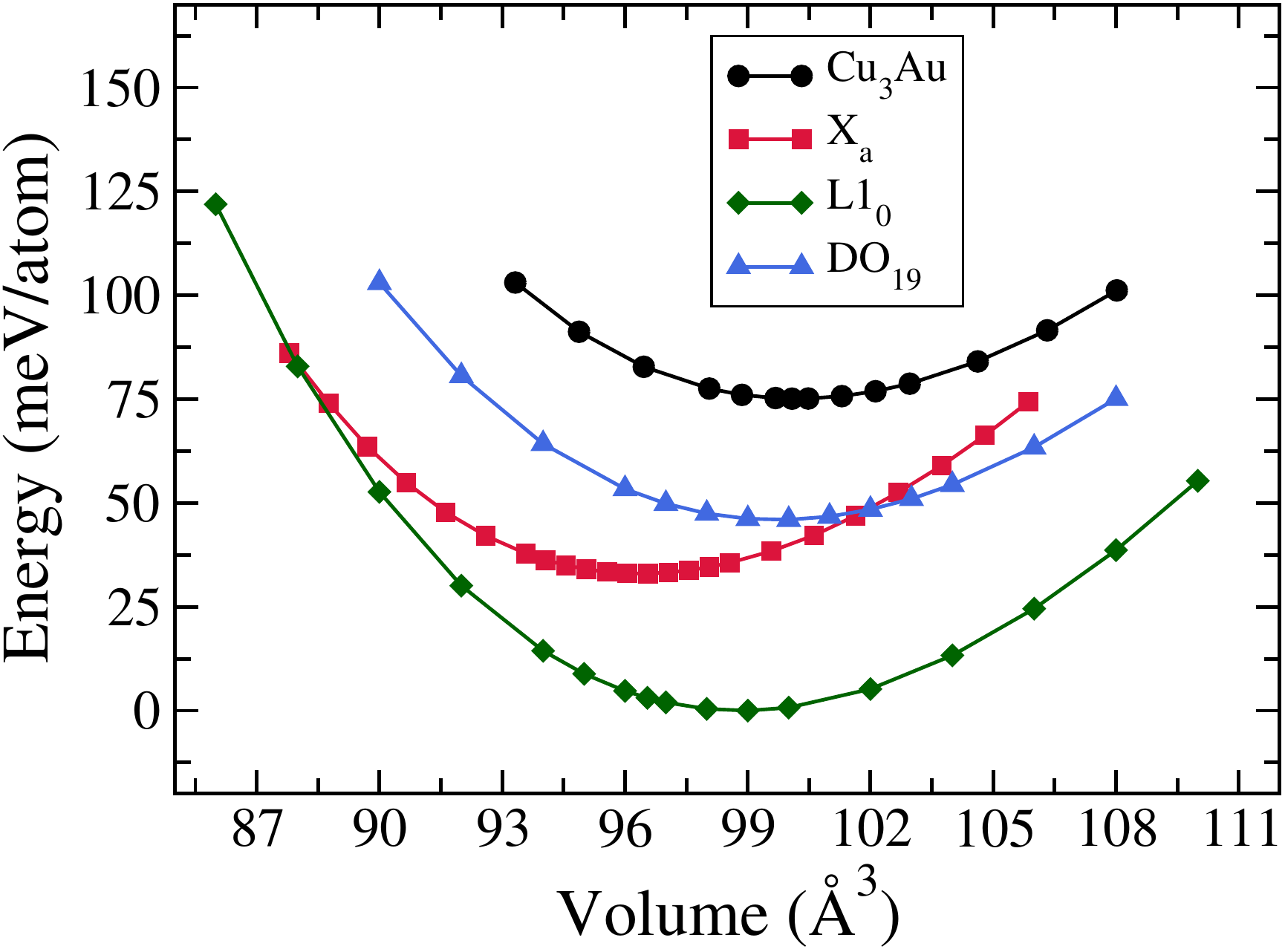,width=0.4\textwidth}\hfill}
\caption{Total energy  as a function of unit cell volume for different structural phases of Mn$_{2}$FeGa.}
\label{fig5}
\end{figure}

Fig. \ref{fig5} shows the variations in the total energies with volume for the four phases of Mn$_{2}$FeGa. We relaxed all the structural parameters to obtain the most stable structure in a given phases. The tetragonal phase comes out to be energetically the most stable phase while the Cu$_{3}$Au phase is having the highest total energy. The hexagonal, and the X$_{a}$ phases lie in between with the X$_{a}$ phase having a total energy in between the hexagonal and the tetragonal phase. The relative energies of these phases obtained this way agrees with the experiments in a way as the tetragonal phase is observed to be the low temperature phase~\cite{GasiAPL13} while annealing to a high temperature stabilises the Cu$_{3}$Au phase; the hexagonal phase being obtained only when the sample is annealed at temperatures in between the annealing temperatures to obtain the L1$_{0}$ and the Cu$_{3}$Au phases. The other interesting feature of the relative energetics is that the DO$_{19}$ and the X$_{a}$ phases are energetically very close near the equilibrium volume of the DO$_{19}$ phase. Since there is no possibility of a direct DO$_{19}$ to L1$_{0}$ transformation due to symmetry constraints~\cite{ZhangJPCM13}, and that a transformation between the X$_{a}$ and DO$_{19}$ is possible, as illustrated in case of Mn$_{3}$Ga~\cite{ZhangJPCM13}, our results imply that the phase transformation from DO$_{19}$ to L1$_{0}$ can take place via the X$_{a}$ phase. Thus, although the X$_{a}$ phase has not yet been obtained experimentally, possibly due to requirement of growth by non-equilibrium methods as it appears to be metastable (Fig. \ref{fig3}), it's presence is very important to understand the stabilities of the DO$_{19}$ and the L1$_{0}$ phases and the possible path of transformation between the two phases.

\begin{figure}[t]
\centerline{\hfill
\psfig{file=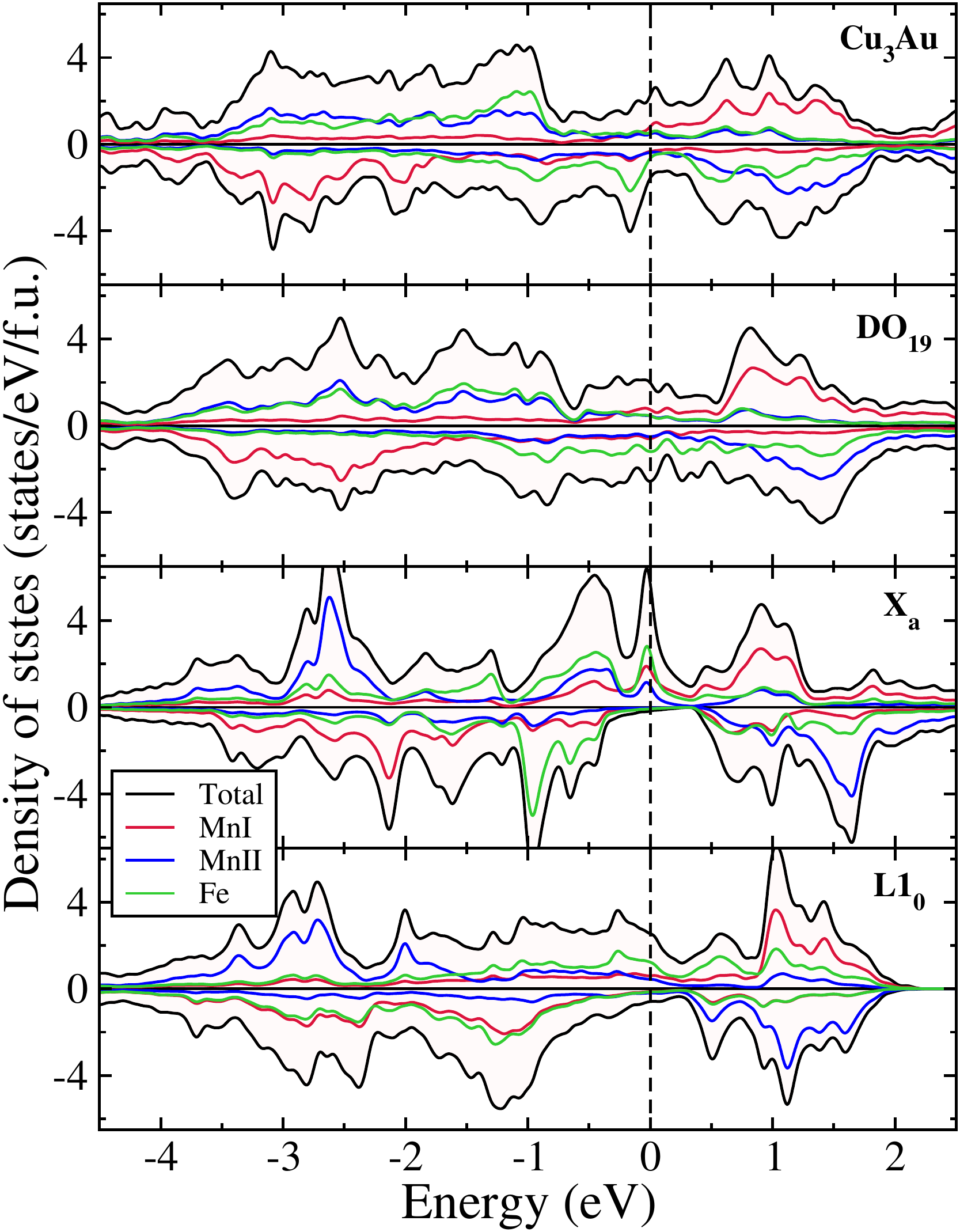,width=0.4\textwidth}\hfill}
\caption{Spin polarised total and atom-projected densities of states of Mn$_{2}$FeGa in various structural phases.}
\label{fig6}
\end{figure}

In order to gain further insight into the stabilities of these phases in Mn$_{2}$FeGa, we take a look at their electronic structures. In Fig. \ref{fig6}, we show the total and atom-projected  densities of states of Mn$_{2}$FeGa in various structural phases. These also help us understanding the behaviour of atomic magnetic moments in each phase. Across all the structures, we find that the minority(majority) band of MnI is nearly full(nearly empty) while the opposite is true for MnII bands. This explains the anti-parallel orientation of the moments associated with the two Mn atoms. For Cu$_{3}$Au and DO$_{19}$ phases, there is strong hybridisation and delocalisation of the MnII and Fe bands in the majority spin channel, while they are more localised in the X$_{a}$ and L1$_{0}$ phases. In the minority spin channel, the hybridisations between MnI and Fe bands are minimal for Cu$_{3}$Au, DO$_{19}$ and X$_{a}$ phases while it is substantial in the L1$_{0}$ phase. In spite of less hybridisation, both MnI and Fe minority states in the Cu$_{3}$Au and DO$_{19}$ phases are delocalised while they are more localised in the X$_{a}$ phase. Thus, the densities of states in the X$_{a}$ phase stand out in contrast to the other three phases; the highlights being more localised atomic densities of states and thus more structured total densities of states in comparison to relatively structureless total densities of states  in other phases. The other major difference observed in the densities of states in the X$_{a}$ phase is a sharp peak at the Fermi level in the majority band and a half-metallic gap in the minority band. The peak at the fermi level is due to hybridisation of MnI and Fe $t_{2g}$ states while the gap in the minority band is flanked by the well separated $t_{2g}$ and the $e_{g}$ states of mainly Fe atom. The peak in the majority band densities of states  implies instability in this structure which triggers the tetragonal distortion and subsequent stabilisation of the L1$_{0}$ phase. The $t_{2g}$ states at the Fermi level in the X$_{a}$ phase now splits into three non-degenerate states due to the lowering of the symmetry, thus washing away the peak. The symmetry lowering gets rid of the half-metallic gap too as the system stabilises by filling in the gap, that is by re-distributing the states into the lower energy states. The delocalisation and strong hybridisation of the MnI and Fe minority states in L1$_{0}$ structure is the outcome of this. The quenching of the Fe moment in the X$_{a}$ structure can also be understood from the electronic structure. The nearly full Fe states in both spin channels reduces the exchange splitting associated with Fe. When the system tetragonally distorts towards the L1$_{0}$ structure, the Fe states fill out significant part of the occupied region in the minority band, while the spectral weight associated with the Fe states in the majority band shifts towards the unoccupied region, thus recovering the exchange splitting observed in the Cu$_{3}$Au and DO$_{19}$ phases. The MnII states in the majority bands associated with the Cu$_{3}$Au and DO$_{19}$ structures are distributed in a nearly identical way, providing an explanation to the almost equal MnII moments in these two phases. The proximity of the MnI moments in these two phases also emerge due to the near identical minority spin densities of states of MnI which implies nearly same exchange splitting associated with the MnI atoms. The Fe moments are also nearly equal in these phases which can be explained in a similar way as Fe hybridises with both Mn in relevant spin channels. The moments associated with MnII in these two phases are less than that of MnI. The relatively less exchange splitting associated with the MnII atoms can be understood as the effect of stronger hybridisation with the Fe states. In the X$_{a}$ and L1$_{0}$ states, the exchange splitting of MnII is more than that of MnI. This happens due to appearance of more localised MnII and Fe states in the majority spin band, thus reducing hybridisations. 

The densities of states imply that Cu$_{3}$Au, DO$_{19}$ and L1$_{0}$ phases are stable as was observed experimentally. The strong hybridisations between Mn and Fe states in either spin channel are found to be responsible behind it. The reduced hybridisations can be correlated with the instability in the X$_{a}$ phase as implied by the large densities of states at the Fermi level. In order to understand the origin of this instability and to predict whether this phase can ever be synthesised, we first check the mechanical stability by computing the elastic constants. The elastic constants for all four phases are given in Table \ref{table3}.

\begin{table}[h]
\centering
\caption{\label{table3} The calculated elastic constants (in GPa) of Mn$_{2}$FeGa in Cu$_{3}$Au, X$_{a}$, L1$_{0}$ and DO$_{19}$ phases.}
\resizebox{0.47\textwidth}{!}{%
\begin{tabular}{c c c c c c c c c}
\hline\hline
Structure  & B &  C$_{11} $ & C$_{12} $ & C$_{13} $ & C$_{33} $ & C$_{44} $ & C$_{66} $ & C$^{\prime}$\\ \hline
Cu$_{3}$Au  & 131.3 & 141.6  & 126.1 & - & - & 33.9 & - &7.8\\
X$_{a}$   &  147.2 & 166.8  & 137.4 & - & - & 127.3 & - &14.7\\
L1$_{0}$  & 140.2 &  292.2 & 0.56 & 124.2 & 179.5 & 100.2 & 11.6 & 145.8\\
DO$_{19}$ & 126.3 &  210.8 & 40.8 & 27.2 & 224.4 & 55.2 & 84.9 & 85\\
\hline\hline
\end{tabular}
}
\end{table}
The calculated elastic constants satisfy all the stability criteria for Cu$_{3}$Au, X$_{a}$, DO$_{19}$ and L1$_{0}$ phases~\cite{Vitosbook07,OzdemirJAC10}. Among all, the shear modulus(C$^{\prime}=(C_{11}-C_{12})/2$) gives an insight to the stability of all the structures with respect to the shear(for hexagonal, $C_{66}=C^{\prime}$). For all the structures, $C^{\prime}$ is positive and indicate the relative mechanical stability of the structure against shear. The $C^{\prime}$ value of tetragonal structure is higher than other structures which might indicate that it is the most stable structure against shear instability. For the X$_{a}$ structure, however, the value of C$^{\prime}$ is sufficiently small, in comparison to C$_{44}$. The elastic anisotropy ratio $A\left(=C^{44}/C^{\prime} \right)$ is a measure of the stability of a crystal with cubic symmetry against stress along $(110)$ planes~\cite{ZenerPRB47}. In the X$_{a}$ phase of Mn$_{2}$FeGa, this ratio is 8.66, indicating that the system can be unstable against anisotropic stress. In order to gain further insight into the possible consequences of it, and thus understand the mechanism of the X$_{a}$ to L1$_{0}$ transformation in this system, we have computed the phonon dispersion relations in both X$_{a}$ and L1$_{0}$ phases (Figs. 8 and 9, supplementary material).  The dispersion relations indicate dynamical stability in both phases. None of the acoustic modes in the X$_{a}$ phase were imaginary, or produced any pronounced softening. These indicate that the transformation from X$_{a}$ to L1$_{0}$ phase is not phonon driven as was the case for Mn$_{2}$NiGa~\cite{PaulJPCM15}. Thus, the phase transition from X$_{a}$ to L1$_{0}$  can be considered due to the well known Jahn-Teller effect.

\subsection{Magnetic Exchange interactions}
\begin{figure}[t]
\centerline{\hfill
\psfig{file=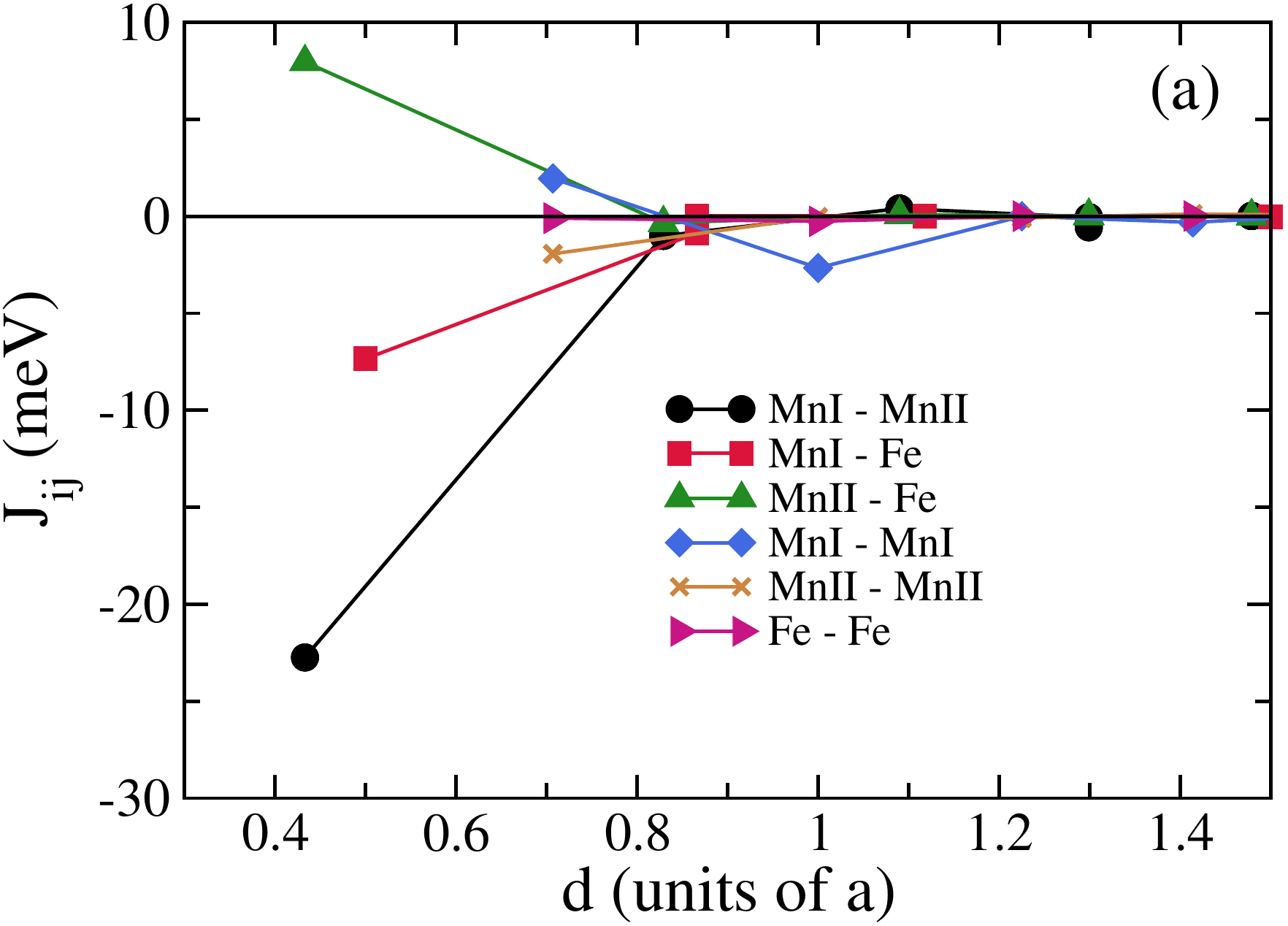,width=0.40\textwidth}\hfill}
\vspace{0.5cm}
\centerline{\hfill
\psfig{file=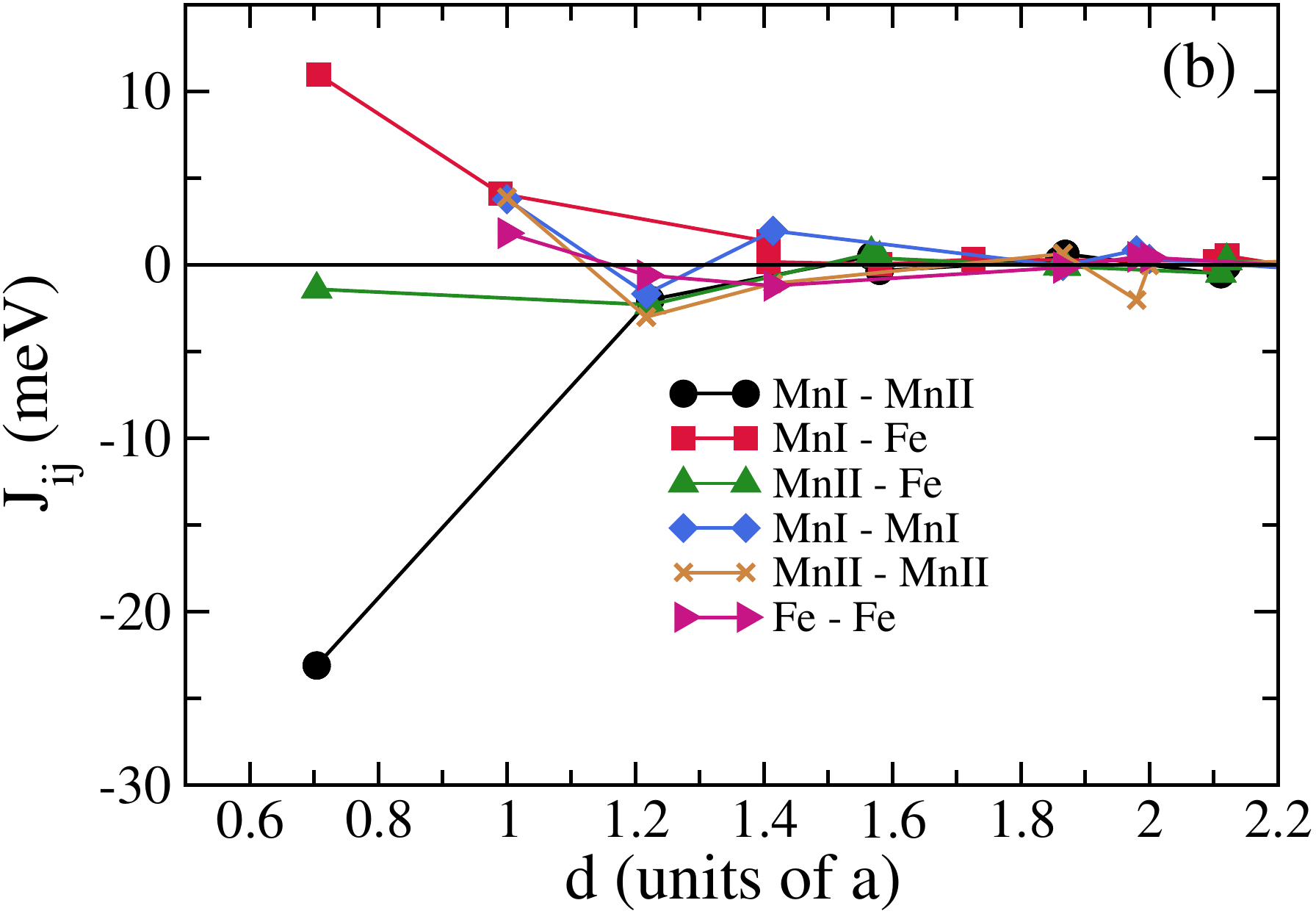,width=0.40\textwidth}\hfill}
\caption{Magnetic exchange interaction(J$_{ij}$) of Mn$_{2}$FeGa as a function of interatomic distance(d)  in (a) X$_{a}$  and (b) L1$_{0}$ phases.}
\label{fig7}
\end{figure}

In order to understand the origin of the magnetic structures in various phases as depicted in sub-section III-A, we now look at the results of various inter-atomic and intra-atomic exchange interactions for the X$_{a}$, L1$_{0}$ and DO$_{19}$ phases. Since the L1$_{0}$ and DO$_{19}$ phases are the most promising ones from applications point of view and the X$_{a}$ phase has strong connection with the L1$_{0}$ phase as has been established in the previous sub-sections, we have investigated the exchange interactions for these three phases. The results for the X$_{a}$ and the L1$_{0}$ phases are presented in Fig. \ref{fig7}(a) and (b), respectively.  For both phases, the nearest neighbour strong antiferromagnetic and ferromagnetic interactions between two different pairs of atoms dominate the scene. The higher neighbour interactions are too small to influence the magnetic structure, which would be collinear. In case of the tetragonal DO$_{22}$ phase of Mn$_{3}$Ga, neutron diffraction experiments~\cite{RodePRB13} found a small canting of Mn spins occupying the 2b sites (the sites occupied by Fe in Fig. \ref{fig2} (b)). The calculated exchange parameters established frustration associated with that particular Mn site, which together with the in-plane magnetic anisotropy at that site, was used to explain the spin canting at that site. In the present case, we do not see any such possibility arising out of the qualitative and quantitative nature of the three major inter-atomic parameters associated with MnI-MnII, Fe-MnI and Fe-MnII pairs. For all three of them, at least the first two nearest neighbour exchange parameters are either ferromagnetic or antiferromagnetic, thus ruling out possibility of magnetic frustration. In Both X$_{a}$ and L1$_{0}$ structures, the overwhelmingly dominating interaction is the antiferromagnetic MnI-MnII. Same was the case for DO$_{22}$ Mn$_{3}$Ga. The qualitative difference between Mn$_{2}$FeGa and Mn$_{3}$Ga magnetic interactions is observed in cases of the exchange parameters involving Fe. In the X$_{a}$ phase, the first neighbour antiferromagnetic Fe-MnI exchange parameter is of the same magnitude as that of the first neighbour ferromagnetic Fe-MnII one although the former pair is at a larger distance. In the L1$_{0}$ phase, we observe a role reversal of these two interactions: the Fe-MnI is now strongly ferromagnetic in the first neighbour, the interaction being ferromagnetic even for higher neighbours, while the Fe-MnII exchange is now weakly antiferromagnetic. This role reversal can be explained from the structural deformation of X$_{a}$ that leads to the L1$_{0}$. The contraction in the basal plane now reduces the Fe-MnI distances along with a subsequent increase in the Fe-MnII distance due to elongation in the [001] direction. This increase in the Fe-MnII distances weaken the exchange interactions between this pair and makes it weakly antiferromagnetic, while the opposite happens for the Fe-MnI pairs which are now closer. The presence of strong ferromagnetic and antiferromagnetic components in the first neighbour exchange interactions of Mn$_{2}$FeGa in the X$_{a}$ and L1$_{0}$ phases along with weak higher neighbour interactions are the reasons behind the collinear ferrimagnetic structure in these phases. Thus, the presence of Fe in the system inflicts a significant ferromagnetic component in the magnetic interactions, in comparison to Mn$_{3}$Ga.   

\begin{figure}[t]
\centerline{\hfill
\psfig{file=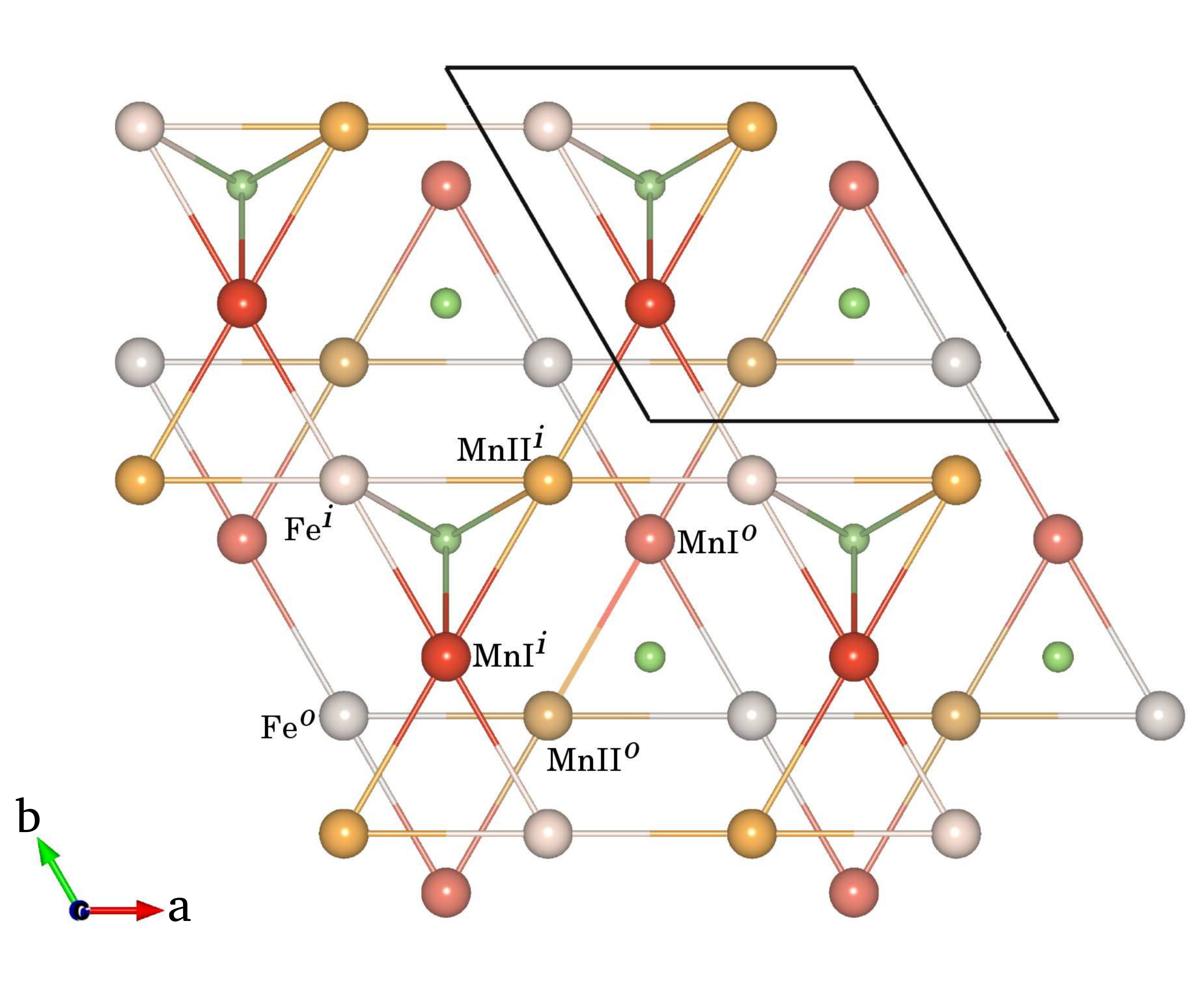,width=0.40\textwidth}\hfill}
\caption{ The nearest neighbour Mn-Mn and Mn-Fe bonds in DO$_{19}$ structure of Mn$_{2}$FeGa are shown(a view from c axis). Network with two adjacent basal planes(the unit is shown on the top right corner with black border) are shown. Atoms marked `\textit{i}' and `\textit{o}'  correspond to whether they are in plane or out of plane, respectively. The relatively darker circles associated with each colour are at the top kagome plane and the relatively lighter circles are at the bottom kagome plane. The colour codes are the ones used in other figures.}
\label{fig8}
\end{figure}

\begin{figure}[t]
\centerline{\hfill
\psfig{file=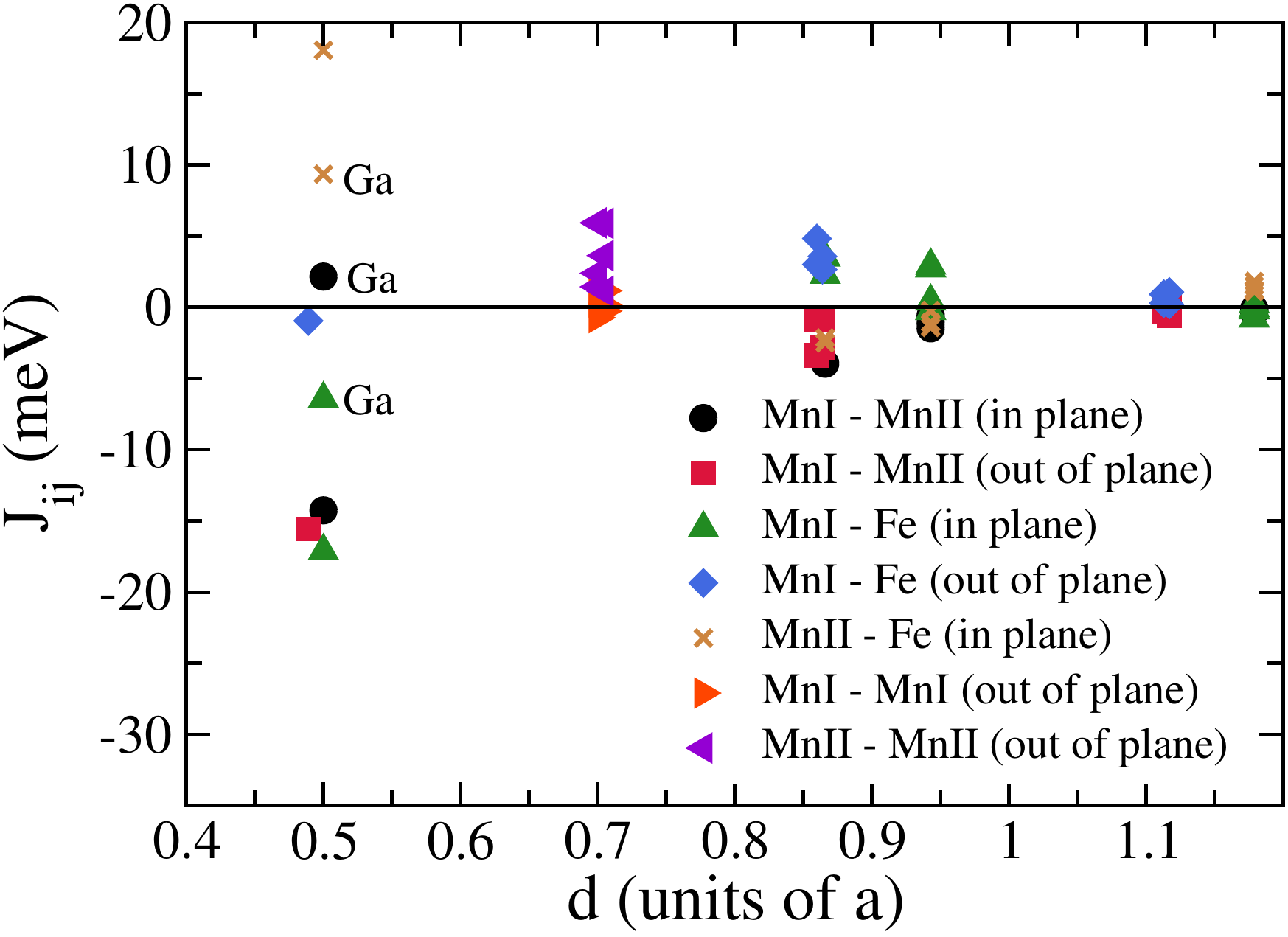,width=0.40\textwidth}\hfill}
\caption{Magnetic exchange interaction(J$_{ij}$) of Mn$_{2}$FeGa  as a function of inter atomic distance(d) in the DO$_{19}$ phase.}
\label{fig9}
\end{figure}

The collinear-like magnetic configuration obtained in the DO$_{19}$ phase of Mn$_{2}$FeGa, as opposed to the non-collinear magnetic structure in Mn$_{3}$Ga where the neighbouring atoms(both in-plane and out-of plane) align themselves in a triangular structure with an angle of 120$^{0}$ between them indicates that Fe plays a vital role. The substantial magnetisation obtained in the measurements~\cite{LiuAPL16} on the DO$_{19}$ phase of Mn$_{2}$FeGa was connected with the possibility of creation of Fe-Mn ferromagnetic matrix in the system. The magnetic exchange interactions calculated for this phase of Mn$_{2}$FeGa and presented in Fig. \ref{fig9} helps understanding the origin of the collinear like magnetic structure obtained from the total energy calculations. The network of the nearest neighbour magnetic atoms in the two neighbouring planes in the DO$_{19}$ structure is shown in Fig. \ref{fig8}. It is clear from the structure that there are two types of triangular networks in the basal planes-one with the shared Ga neighbours and one without. Consequently, the strengths of the interactions between the same pair of specie at the same distance would be different in two networks. This was found out to be the case for Mn$_{3}$Ga where the in-plane interaction between Mn atoms with shared Ga were comparable to the interaction without shared Ga, which ensured the 120$^{0}$ orientation between the Mn neighbours belonging to different networks~\cite{KhmelevskyiPRB16}. In the present case of Mn$_{2}$FeGa, we find that the in plane exchange parameters for the three pairs of magnetic atoms are heterogeneous. While the MnI-MnII and MnI-Fe first neighbour interactions without shared Ga are strongly antiferromagnetic, the MnII-Fe interactions are strongly ferromagnetic with magnitudes of the three interactions comparable. The same qualitative feature are  observed for the interactions with networks having Ga atoms in between, although the strength of the interactions are reduced, yet non-negligible. These ensure that the frustration in the system is destroyed and we obtain a collinear like magnetic structure. The out-of-planes interactions too fail to induce any frustration as the dominating second neighbour out-of-plane ferromagnetic interactions are about 3 times smaller than the first neighbour out of plane antiferromagnetic interaction. Thus, the presence of Fe in the system indeed introduces substantial Fe-Mn ferromagnetic interactions in the system getting rid of the possible frustration as was conjectured~\cite{LiuAPL16}.

\section{Conclusions}
The presented investigation of the structural and magnetic properties of four different structural phases of Mn$_{2}$FeGa in the framework of same first-principles method enables us to understand the microscopic physics associated with different physical aspects as well as the similarities and differences with the prototype Mn$_{3}$Ga from which the system under present investigation can be derived, and which is having similar promising applications in the area of spintronics. We find striking similarity between the two systems with regard to relative stabilities of the structural phases. Particularly important was to investigate the importance of the X$_{a}$ phase which has not been synthesised experimentally. We find that the X$_{a}$ phase provides an important connection between the hexagonal DO$_{19}$ and the tetragonal L1$_{0}$ phase. However, we find that the phase is mechanically unstable to anisotropic stress. The origin of the electronic instability associated with this phase appears to be Jahn-Teller effect. The analysis of the electronic structures reveal that the significant hybridisations between Mn and Fe atoms are responsible for the stabilities and atomic magnetism of the other three phases investigated. While Mn$_{2}$FeGa has substantial similarities with Mn$_{3}$Ga in these aspects, there is significant qualitative differences in the magnetic structures. Unlike the tetragonal DO$_{22}$ phase of Mn$_{3}$Ga, where a canted spin structure associated with one of the Mn sites was observed, the tetragonal L1$_{0}$ phase of Mn$_{2}$FeGa has a collinear magnetic structure. The results of magnetic exchange interactions clearly showed that the equally strong ferromagnetic Fe-Mn interactions and antiferromagnetic Mn-Mn interactions are responsible for the collinear ferrimagnetic structure in this phase as well as in the X$_{a}$ phase. The magnetic structure in the hexagonal DO$_{19}$ phase is where we observe substantial impact of the presence of Fe in the system. The non-collinear magnetic structure in Mn$_{3}$Ga, which was an artefact of the geometric frustrations and magnetic frustrations due to Mn-Mn exchange interactions now gives way to a collinear like magnetic structure with spins confined to be $a-b$ plane, which is energetically almost degenerate to a collinear configuration with $z$-axis as the easy axis. The origin of this is found to be strong ferromagnetic Fe-Mn components in an antiferromagnetic host as was suggested in the experiment~\cite{LiuAPL16}.

\section*{ACKNOWLEDGMENT}
Computational support from C-DAC, Pune, India, PARAM-ISHAN Supercomputer, IIT Guwahati and Department of Physics, IIT Guwahati, India through DST-FIST programme are gratefully acknowledged.


\begin{thebibliography}{10}

\bibitem{WinterlikAM12}
J.~Winterlik et~al.,
\newblock Advanced Materials {\bf 24}, 6283 (2012).

\bibitem{SinghPRL13}
S.~Singh et~al.,
\newblock Phys. Rev. Lett. {\bf 109}, 246601 (2012).

\bibitem{KubotaAPL09}
T.~Kubota, S.~Tsunegi, M.~Oogane, S.~Mizukami, T.~Miyazaki, H.~Naganuma, and
  Y.~Ando,
\newblock Appl. Phys. Lett. {\bf 94} (2009).

\bibitem{OuardiPRL13}
S.~Ouardi, G.~H. Fecher, C.~Felser, and J.~K\"ubler,
\newblock Phys. Rev. Lett. {\bf 110}, 100401 (2013).

\bibitem{LiuAPL05}
G.~Liu et~al.,
\newblock Appl Phys Lett {\bf 87}, 262504 (2005).

\bibitem{BarmanPRB08}
S.~Barman and A.~Chakrabarti,
\newblock Physical Review B {\bf 77}, 176401 (2008).

\bibitem{PaulJAP11}
S.~Paul and S.~Ghosh,
\newblock Journal of Applied Physics {\bf 110}, 063523 (2011).

\bibitem{MeshcheriakovaPRL14}
O.~Meshcheriakova et~al.,
\newblock Physical review letters {\bf 113}, 087203 (2014).

\bibitem{WinterlikPRB11}
J.~Winterlik et~al.,
\newblock Physical Review B {\bf 83}, 174448 (2011).

\bibitem{GasiAPL13}
T.~Gasi, A.~K. Nayak, J.~Winterlik, V.~Ksenofontov, P.~Adler, M.~Nicklas, and
  C.~Felser,
\newblock Applied Physics Letters {\bf 102}, 202402 (2013).

\bibitem{WollmannPRB14}
L.~Wollmann, S.~Chadov, J.~K\"ubler, and C.~Felser,
\newblock Phys. Rev. B {\bf 90}, 214420 (2014).

\bibitem{WollmannPRB15}
L.~Wollmann, S.~Chadov, J.~K\"ubler, and C.~Felser,
\newblock Phys. Rev. B {\bf 92}, 064417 (2015).

\bibitem{NayakPRL13}
A.~Nayak, M.~Nicklas, S.~Chadov, C.~Shekhar, Y.~Skourski, J.~Winterlik, and
  C.~Felser,
\newblock Physical review letters {\bf 110}, 127204 (2013).

\bibitem{Kundumodulation17}
A.~Kundu, M.~E. Gruner, M.~Siewart, A.~Hucht, P.~Entel, and S.~Ghosh,
\newblock arXiv preprint arXiv:1703.06705  (2017).

\bibitem{KurtPSS11}
H.~Kurt, K.~Rode, M.~Venkatesan, P.~Stamenov, and J.~Coey,
\newblock physica status solidi (b) {\bf 248}, 2338 (2011).

\bibitem{RodePRB13}
K.~Rode et~al.,
\newblock Physical Review B {\bf 87}, 184429 (2013).

\bibitem{KharelJPCM14}
P.~Kharel, Y.~Huh, N.~Al-Aqtash, V.~Shah, R.~F. Sabirianov, R.~Skomski, and
  D.~J. Sellmyer,
\newblock Journal of Physics: Condensed Matter {\bf 26}, 126001 (2014).

\bibitem{ZhangJPCM13}
D.~Zhang, B.~Yan, S.-C. Wu, J.~K{\"u}bler, G.~Kreiner, S.~S. Parkin, and
  C.~Felser,
\newblock Journal of Physics: Condensed Matter {\bf 25}, 206006 (2013).

\bibitem{KhmelevskyiPRB16}
S.~Khmelevskyi, A.~V. Ruban, and P.~Mohn,
\newblock Phys. Rev. B {\bf 93}, 184404 (2016).

\bibitem{PaulJPCM15}
S.~Paul, B.~Sanyal, and S.~Ghosh,
\newblock J. Phys.: Condens. Matter. {\bf 27}, 035401 (2015).

\bibitem{KurtAPL12}
H.~Kurt, K.~Rode, H.~Tokuc, P.~Stamenov, M.~Venkatesan, and J.~Coey,
\newblock Applied Physics Letters {\bf 101}, 232402 (2012).

\bibitem{KurtPRB11}
H.~Kurt, K.~Rode, M.~Venkatesan, P.~Stamenov, and J.~M.~D. Coey,
\newblock Physical Review B {\bf 83}, 020405(R) (2011).

\bibitem{NayakNM15}
A.~K. Nayak et~al.,
\newblock Nature materials {\bf 14}, 679 (2015).

\bibitem{NiesenIEEE16}
A.~Niesen, C.~Sterwerf, M.~Glas, J.~M. Schmalhorst, and G.~Reiss,
\newblock IEEE Transactions on Magnetics {\bf 52}, 1 (2016).

\bibitem{BettoarXiv17}
D.~Betto et~al.,
\newblock arXiv preprint arXiv:1704.01326  (2017).

\bibitem{KalachearXiv17}
A.~Kalache, A.~Markou, S.~Selle, T.~H{\"o}che, G.~H. Fecher, and C.~Felser,
\newblock arXiv preprint arXiv:1705.10668  (2017).

\bibitem{LiuAPL16}
Z.~Liu, Y.~Zhang, H.~Zhang, X.~Zhang, and X.~Ma,
\newblock Applied Physics Letters {\bf 109}, 032408 (2016).

\bibitem{PAW94}
P.~E. Bl\"ochl,
\newblock Phys. Rev. B {\bf 50}, 17953 (1994).

\bibitem{VASP196}
G.~Kresse and J.~Furthm\"uller,
\newblock Phys. Rev. B {\bf 54}, 11169 (1996).

\bibitem{VASP299}
G.~Kresse and D.~Joubert,
\newblock Phys. Rev. B {\bf 59}, 1758 (1999).

\bibitem{PBEGGA96}
J.~P. Perdew, K.~Burke, and M.~Ernzerhof,
\newblock Phys. Rev. Lett. {\bf 77}, 3865 (1996).

\bibitem{MP89}
M.~Methfessel and A.~T. Paxton,
\newblock Phys. Rev. B {\bf 40}, 3616 (1989).

\bibitem{Vitosbook07}
L.~Vitos,
\newblock {\em Computational quantum mechanics for materials engineers: the
  EMTO method and applications},
\newblock Springer Science \& Business Media, 2007.

\bibitem{EbertRPP11}
H.~Ebert, D.~Koedderitzsch, and J.~Minar,
\newblock Rep. Prog. Phys. {\bf 74}, 096501 (2011).

\bibitem{LiechtensteinJMMM87}
A.~Liechtenstein, M.~Katsnelson, V.~Antropov, and V.~Gubanov,
\newblock J. Magn. Magn. Mater. {\bf 67}, 65  (1987).

\bibitem{LuoJAP08}
H.~Luo et~al.,
\newblock Journal of Applied Physics {\bf 103}, 083908 (2008).

\bibitem{krenSSC70}
E.~Kr{\'e}n and G.~K{\'a}d{\'a}r,
\newblock Solid State Communications {\bf 8}, 1653 (1970).

\bibitem{OzdemirJAC10}
S.~O. Kart and T.~Cagın,
\newblock Journal of Alloys and Compounds {\bf 508}, 177  (2010).

\bibitem{ZenerPRB47}
C.~Zener,
\newblock Phys. Rev. {\bf 71}, 846 (1947).

\end{thebibliography}
\end{document}